\documentclass[letterpaper,twocolumn,amsmath,amssymb,prb]{revtex4-2}
\usepackage{siunitx}
\usepackage{graphicx}
\usepackage{float}
\usepackage{booktabs}
\usepackage{setspace}
\usepackage[svgnames]{xcolor}
\usepackage[unicode,colorlinks=true,citecolor=blue,linkcolor=blue,urlcolor=NavyBlue]{hyperref}
\setcitestyle{numbers,comma,sort&compress}
\usepackage{url}
\usepackage[normalem]{ulem}
\usepackage[nameinlink,capitalize]{cleveref}
\usepackage{mhchem}
\begin{document}
\title{Evidence for room temperature superconductivity associated with a first-order phase transition}
\author{N. Zen\vspace{1ex}}
\affiliation{National Institute of Advanced Industrial Science and Technology,\\Tsukuba Central 2-10, Ibaraki 305-8568, Japan}
\begin{abstract}
By making periodic thru-holes in a suspended film, the phonon system can be modified. Motivated by the BCS theory, the technique---so-called phonon engineering---was applied to a metallic niobium sheet. It was found that its electrical resistance dropped to zero at 175 K, and the zero-resistance state persisted up to 290 K in the subsequent warming process. Despite the initial motivation, neither these high transition temperatures nor the phase transition with thermal hysteresis can be accounted for by the BCS theory. Therefore, we abandon the BCS theory. Instead, it turns out that the metallic holey sheet is partly oxidized to form a niobium--oxygen square lattice, which has points of resemblance to a copper--oxygen plane, the fundamental component of cuprate high-$T_{c}$ superconductors. Therefore, the pairing mechanism underlying this study should be related to that of cuprate high-$T_{c}$ superconductors, which we may not yet understand. In addition to the electrical results of zero resistance, the holey sheet exhibited a decrease in magnetization upon cooling, i.e., the Meissner effect. Moreover, the remnant magnetization was clearly detected at 300 K, which can only be attributed to persistent currents flowing in a superconducting sample. Thus, this study meets the established criteria for a conclusive demonstration of true superconductivity. Finally, the superconducting transition with the unambiguous thermal hysteresis is discussed. According to Halperin, Lubensky, and Ma, or HLM for short, any superconducting transition must \emph{always} be first order with thermal hysteresis because of the intrinsic fluctuating magnetic field. The HLM theory is very compatible with the highly oriented system harboring two-dimensional superconductivity.
{}
\end{abstract}
\maketitle
\section{Introduction}
\label{sec:intro}
\begin{figure}[b]
\centering
\includegraphics[width=66mm]{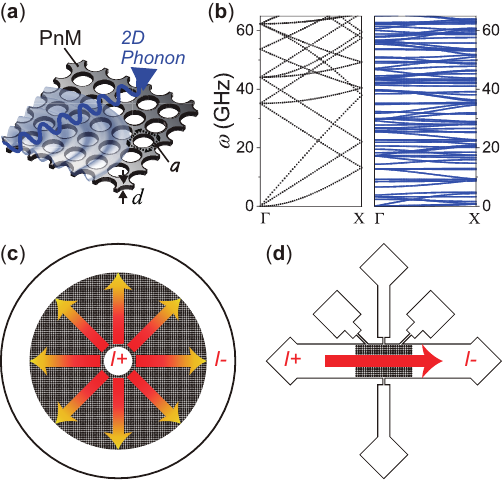}
\caption{\textbf{Phonon engineered metallic system (PnM).} (\textbf{a}) Schematic representation of a 2D PnM. The thickness and lattice constant are indicated by $d$ and $a$, respectively. (\textbf{b}) Phonon dispersions of an Nb sheet with $d$ of 150 nm. (Left) Without perforation. (Right) With periodic perforation designed as a PnM with $a$ of 20 $\si{\micro m}$. (\textbf{c}) and (\textbf{d}) Schematic design of the PnM sample used for the previous study~\cite{Zen2019} and for this study, respectively.
\label{fig1}}
\end{figure}

According to the BCS theory of superconductivity~\cite{BCS1957}, coherent phonons play a key role in coupling electrons. In other words, a critical temperature ($T_{c}$) of superconducting transition is closely associated with the property of coherent phonons, namely, the phonon dispersion relation. It is generally supposed that the phonon dispersion is material dependent and hence cannot be modified. But it becomes possible, simply by drilling periodic thru-holes in a material, as shown in \cref{fig1}(a). The periodic perforation induces the elastic analogue of Bragg's interference and therefore affects the coherence of phonon propagation. The technique, known as phonon engineering~\cite{Zen2014}, has attracted physicists and material scientists' attention because of its ability to control thermal properties of dielectric materials~\cite{Mal2015,Soto2019,Nom2020,JAP2021,Pekola2021}. On the other hand, another possibility of the phonon engineering---whether an artificially modified phonon system affects the electron system or not---is not yet getting attention.

In the previous study~\cite{Zen2019}, the phonon engineering was applied to a metallic system for the first time. The used material was pure niobium (Nb), the well known conventional superconductor with $T_{c}$ of 9 K. Expecting a change in the $T_{c}$, hopefully to be increased from 9 K to 20's K, the upper limit of $T_{c}$ for a phonon-engineered Nb according to the BCS-McMillan theory of superconductivity~\cite{Mc1968}, a pure Nb film with a thickness ($d$) of 150 nm was periodically perforated to form a two-dimensional (2D) square lattice with a lattice constant ($a$) of 20 $\si{\micro m}$. The calculated phonon dispersion of the engineered Nb sheet using these $d$ and $a$ is shown on the right panel of \cref{fig1}(b). The left panel shows that of an Nb sheet with the same $d$ without perforation. Obviously, overall phonon bands are forcibly flattened by the phonon engineering, hence, the change in the $T_{c}$ could be expected. Despite the expectation, however, the $T_{c}$ neither increased nor decreased. Instead, the engineered Nb sheet underwent a metal--insulator transition at 43 K~\cite{Zen2019}. Independently performed resistive and magnetic measurements revealed that the metal--insulator transition was caused by the Anderson localization of the electron system.

The Anderson localization is the indication of spatially disordered charge distribution in the sample. \Cref{fig1}(c) is a schematic design of the previous sample. Its Corbino disk shape, wherein an excitation current for electrical measurements flows through the sample radially from the center to the periphery reducing its own current density, may cause a disordered charge distribution and the consequent metal--insulator transition.

In this study, a rectangular geometry is applied to the phonon-engineered metallic system (PnM) instead of the Corbino geometry. As shown in \cref{fig1}(d), an excitation current uniformly flows in a single direction keeping a constant current density, therefore, the Anderson localization would be suppressed. Then we can see the ending of the tiny spark of reason inspired by the BCS theory---whether a simple modification of phonon dispersions affects a $T_{c}$ of conventional superconductivity or not.

\section{Material and Methods}
\label{sec:matmet}
\subsection{Sample fabrication}
\label{sec:samfab}
First, a silicon dioxide (SiO\textsubscript{2}) sacrificial layer of 1.0-$\si{\micro m}$ thickness was deposited on a p-type silicon (Si) wafer having the thickness, diameter and orientation, respectively, 0.4 mm, 3 inch ($\approx$ 76 mm) and (100) by chemical vapor deposition (PD-270STL, Samco) with the stage temperature kept at 80 \si{\degreeCelsius}. The pressure of the mixture of gases of TEOS (tetraethoxysilane) and O\textsubscript{2} was 30 Pa, and the total deposition time was 42 minutes. Post deposition, an Nb film of 150-nm thickness was deposited on the SiO\textsubscript{2} layer by sputtering (M12-0130, Science Plus) at 10 \si{\degreeCelsius} using Ar gas at 1.0 Pa for 130 s. Subsequently, an i-line chemical resist (PFi-245, Sumitomo Chemical) was spin-coated to be a thickness of 300 nm on the Nb layer, and the sample patterning was performed to form the square lattice with the lattice constant $a$ of 20 $\si{\micro m}$ using an i-line stepper (NSR-2205i12D, Nikon TEC) with an exposure time of 350 ms. The exposed region of the Nb layer was removed by reactive ion etching (RIE-10NR, Samco) using SF\textsubscript{6} gas at 10.0 Pa for the total etching time of 210 s. After a protective chemical resist (PFi-68A7, Sumitomo Chemical) was spin-coated, the resulting wafer was cut into 5-mm squares using a dicing machine (DAD522, DISCO), and the protective chemical resist was removed. Finally, the SiO\textsubscript{2} sacrificial layer under the already patterned Nb layer was removed by an HF dry etcher (memsstar\textsuperscript{\circledR}SVR\textsuperscript{TM} vHF, Canon). The diced samples were exposed under the mixture of 250-sccm HF gas and water vapor consisting of 100-sccm N\textsubscript{2} and 10-mg H\textsubscript{2}O with the stage temperature kept at 5 \si{\degreeCelsius} for the duration of 120 s and 360 s, respectively, for the 8-Torr step and for the subsequent 9-Torr step. The self-standing PnM-Nb structure was inspected using a laser microscope (LEXT OLS4000, Olympus).

\begin{figure}[h!]
\centering
\includegraphics[width=66mm]{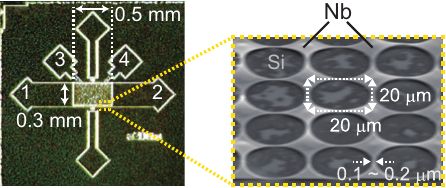}
\caption{Optical micrograph of the sample. Electrical pads are numbered. Right panel, an SEM of the region surrounded by the yellow dotted square. The Nb holey sheet is self standing, approximately 1 $\si{\micro m}$ apart from the Si substrate underneath. The lattice constant of the giant square lattice is 20 $\si{\micro m}$. The narrowest width of the bridge-like constrictions is 0.1--0.2 $\si{\micro m}$.
\label{fig2}}
\end{figure}

\Cref{fig2} shows an optical micrograph of the PnM-Nb sample, and its scanning electron micrograph (SEM) is shown on the right panel. The sample has a rectangle shape with an area of 0.3$\times$0.5 mm\textsuperscript{2}. The Nb sheet with $d$ of 150 nm has the square lattice structure with $a$ of 20 $\si{\micro m}$ and is self standing, approximately 1 $\si{\micro m}$ apart from the Si substrate underneath. The self-standing structure is essential; even if a sample has the periodic structure designed as a PnM, it never shows an anomalous phase transition if it is not self standing. A possible reason for this can be found in the following section.

\subsection{Sample characterization}
\label{sec:samcha}
\begin{figure}[h!]
\centering
\includegraphics[width=0.82\linewidth]{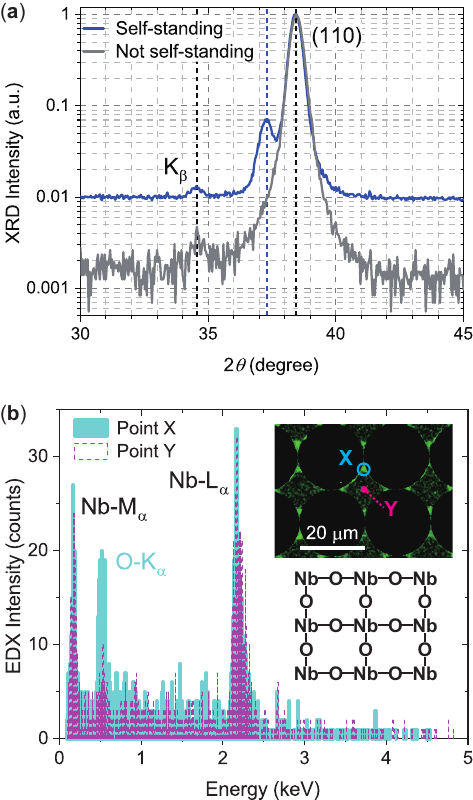}
\caption{{\if0\textbf{Structural properties of the sample.} \fi}(\textbf{a}) XRD spectra of the self-standing sample and not-self-standing one. (\textbf{b}) EDX spectra of the self-standing sample, where the points ``X'' and ``Y'' are indicated in the SEM-EDX result (inset). Bottom of the inset, a schematic configuration of the SEM-EDX result. The XRD and SEM-EDX results were obtained using RINT-Ultima III (Rigaku) and JSM-7200F (JEOL), respectively.
\label{fig3}}
\end{figure}

\Cref{fig3}(a) shows x-ray diffraction (XRD) spectra of the self-standing sample and a not-self-standing one which was prepared by skipping the final SiO\textsubscript{2} removal procedure. Both spectra have main peaks at 38.3 degree which corresponds to the usual Nb atomic lattice spacing of 3.3 \AA. By contrast, only the self-standing sample exhibits an extra peak at 37.3 degree. The extra peak corresponds to the atomic lattice spacing of 3.4 \AA, indicating that the self-standing sample contains anomalous regions where the Nb lattice spacing is expanded, a little wider than the usual 3.3 \AA. The unexpected expansion may be due to the removal of the SiO\textsubscript{2} sacrificial layer under the Nb layer, because the removal releases an in-plane stress of the Nb layer accumulated during the Nb sputtering process.

By the way, it is well known that light elements such as H, C, N, O in the surrounding atmosphere easily invade a metal. The wider the spacing is, the higher amounts invade. \Cref{fig3}(b) shows energy dispersive x-ray (EDX) spectra of the self-standing sample, together with its SEM-EDX result in the inset. The accelerating voltage and probe current were 5 kV and 1 nA, respectively, and the integration time was 30 minutes. Obviously, the region pointed as ``X'' includes higher amounts of oxygen than that of ``Y''. Due to the aforementioned anomalous lattice expansion, the PnM-Nb sample thus forms a niobium--oxygen plane in a square lattice configuration. Except for the lattice constant and the composite metal, the exposed configuration is the same as that of the copper--oxygen plane, the fundamental component of cuprate high-$T_{c}$ superconductors~\cite{Bed1986,Ashburn1987,Muller1987}. The relevance of the exposed configuration to superconductivity will be determined in the future when we better understand the pairing mechanism of cuprate high-$T_{c}$ superconductors.

\subsection{Measurement methods}
\label{sec:meamet}
First, temperature ($T$) dependence of the electrical resistance ($R$) was investigated. The \emph{R--T} measurement was performed using the PPMS (Quantum Design) under zero magnetic field. The sample chips were mounted on a PPMS sample puck using vacuum grease (Apiezon N, M\&I Materials), and the electrical contacts for the sample to the PPMS sample puck were made by aluminum wire bonding. The resistance was measured by the two-probe method using the electrical pads denoted in \cref{fig2}; pads number 1 and 2 were used. The PPMS was operated in the AC drive mode with the standard calibration mode, and the number of readings taken was twenty five. That is, at each temperature, positively and negatively oscillating 8.33-Hz square-wave excitation current with the amplitude of $\pm$10 $\si{\micro A}$ was repeatedly applied to the sample twenty five times, and the output voltage was obtained by averaging output values to minimize the DC offset error; all of these procedures were done by the PPMS automatically. One \emph{R--T} cycle consists of a cooling process from 300 K to 2 K and a subsequent warming process from 2 K to 300 K. The \emph{R--T} cycle (300 K $\rightarrow$ 2 K $\rightarrow$ 300 K) was continuously repeated for eight times.

Second, after finishing the \emph{R--T} measurement, the sample chip was taken out of the PPMS and was installed in the MPMS (Quantum Design), and its temperature ($T$) dependence of the magnetization ($M$) was investigated. The sample chip was positioned in a plastic straw. After the temperature was lowered down to 4.2 K under zero magnetic field, the magnetic field of 1000 Oe was applied, perpendicularly to the sample surface, and then the centering procedure was performed. The magnetization $M$ of the sample chip was scanned by moving the entire straw through the SQUID ring. The oscillation amplitude, frequency and cycles to average were 0.3 cm, 4 Hz and 40 cycles, respectively. The number of scans per measurement was three. While warming the sample chip from 4.2 K to 300 K, the zero field cooling (ZFC) measurement was performed. After that, the field cooling (FC) measurement was subsequently performed while cooling the sample chip from 300 K to 5.1 K with the applied magnetic field of 1000 Oe unchanged.

236 days later, the above \emph{M--T} cycles were applied to the same sample again, and the temperature was raised to 300 K with the applied magnetic field of 1000 Oe unchanged. Subsequently, the applied field was once decreased to 0 Oe, and then applied magnetic field ($H$) dependence of the magnetization ($M$) was investigated at 300 K. The oscillation amplitude, frequency, cycles to average, number of scans were 0.5 cm, 4 Hz, 40 cycles and three, respectively.

Finally, the critical magnetic field ($H_{c}$) at 300 K was investigated using the PPMS. The sample used for this measurement was another PnM-Nb which preserved zero resistance at 300 K during an \emph{R--T} cycle performed in advance. Under various magnitudes of perpendicular magnetic field, current was applied to the sample, and the output voltage was measured by the four-probe method using the electrical pads denoted in \cref{fig2} (pads number 1, 2 for applying current; 3, 4 for measuring voltage). The delta mode (6221/2182A combination, Keithley), which can minimize constant thermoelectric offsets, was externally connected to the sample in the PPMS. The current pulse with a width of 10 ms and a period of 100 ms was applied, and the output voltage was measured in the minimum range of 10 mV.

\section{Results}
\label{sec:res}
\subsection{\emph{R--T} results}
\label{sec:rdrop}
\begin{figure*}
\centering
\includegraphics[width=0.87\textwidth]{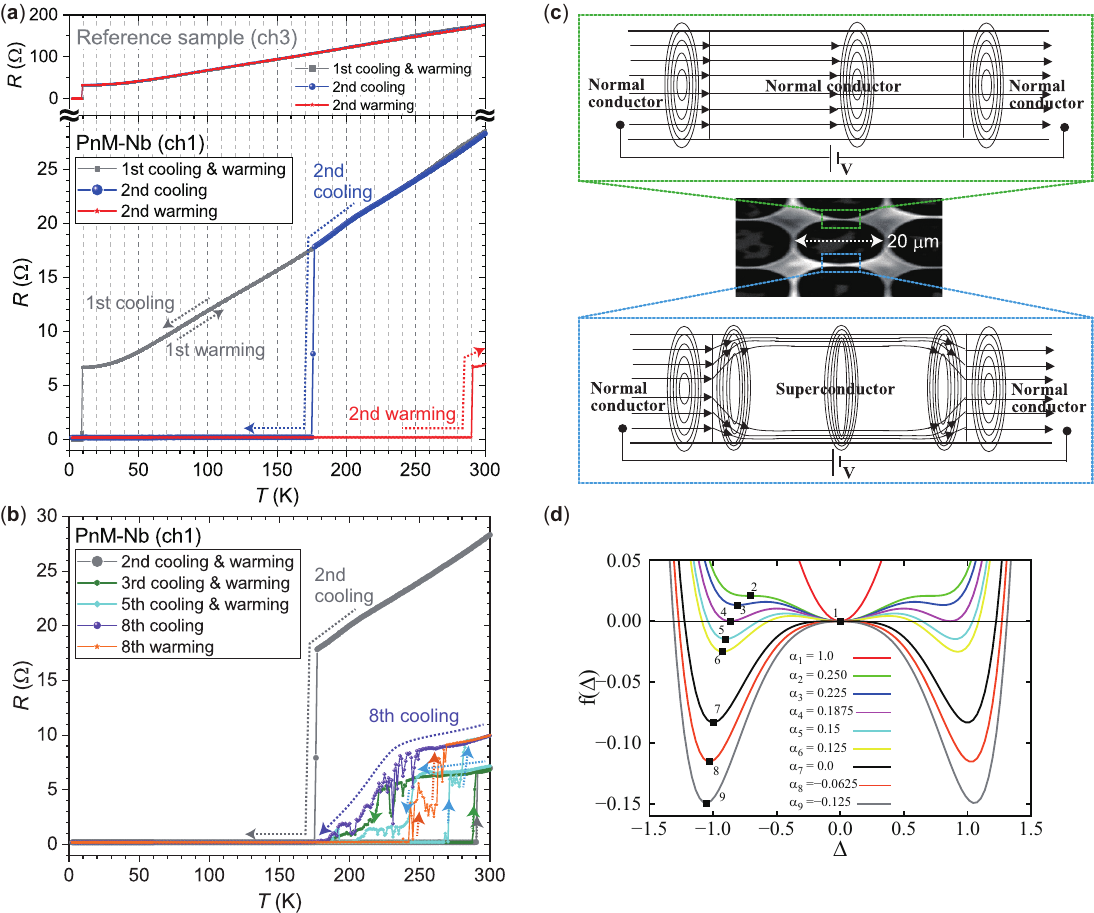}
\caption{\textbf{\emph{R--T} results and possible involvement of magnetic flux for the first order phase transition.} (\textbf{a}) Electrical resistance versus temperature of a reference Nb sample (upper panel) and that of a PnM-Nb sample (bottom panel) using the two-probe method, in the absence of an externally applied magnetic field. They were measured at the same time in the PPMS under the same condition. Only \emph{R--T} results for the first and second cycles are shown. (\textbf{b}) \emph{R--T} results of the PnM-Nb sample for all the rest of eight cycles. (\textbf{c}) Possible involvement of magnetic flux in the superconducting transition for an Nb holey sheet, based on the theory of \emph{hole superconductivity}. Center panel, a portion of an SEM image of the Nb holey sheet; a void with the diameter of approximately 20 $\si{\micro m}$ is surrounded by four narrow bridges. Upper and lower panel, a schematic normal bridge and a superconducting one with current streamlines and magnetic field lines (circles), respectively, from Ref.~\cite{HirschWire2021}. Even in the absence of an externally applied magnetic field, a self field generated by an excitation current for measuring resistance is expelled from a bridge becoming superconducting. As the dynamics of magnetic flux is involved in the superconducting transition, it can be a first order phase transition. This effect is significant for a system wherein magnetic orientation of the expelled flux is not to be averaged out. For details, read text. (\textbf{d}) Reason why warming is \emph{always} beneficial to the zero-resistance state presented in this study, based on the Ginzburg-Landau theory: the key feature of a first-order phase transition. Free energy density as a function of the order parameter, for various temperatures, is dictated by differing choices of $\alpha$, as indicated in the legend, from Ref.~\cite{HMarxiv2021Jan}. The highest temperature is for $\alpha_1 = 1$ and progresses downwards towards $\alpha_9 = -0.125$. Note that hysteresis may arise because, on lowering the temperature from $\alpha_4$ to $\alpha_5$ the system may choose to remain in the $\Delta = 0$ state. Only when the temperature is lowered to $\alpha_7$ (when $\Delta = 0$ is no longer a local minimum), does the system have to transition to a state with non-zero $\Delta$ (we show the negative solutions only as square points for clarity), at which point the resistance would drop to zero. Upon increasing the temperature, a similar phenomenon may occur, whereby the non-zero $\Delta$ state remains, until finally $\alpha_2$ is reached, and then a transition has to occur to the $\Delta=0$ state with finite resistance.
\label{fig4}}
\end{figure*}

\Cref{fig4}(a) shows the temperature ($T$) dependence of electrical resistance ($R$) under zero magnetic field, measured by the two-probe method. The upper panel shows that of a reference Nb sample which was mounted on the PPMS sample puck together with the PnM-Nb sample and was measured at the same time under the same condition. (The PPMS can measure three samples at the same time.) As shown, the reference sample undergoes the superconducting (SC) transition normally at the usual $T_{c}$ for Nb ($\approx$~9~K) both for the first and second temperature cycles. Although there shows only the first and second cycles, the result is the same for all the rest of eight \emph{R--T} cycles. The standard deviation of the onset SC transition temperature for all the eight cycles is 9.0000$\pm$0.0006 K, indicating that during the whole cycles the thermometer of the PPMS was working properly and that the temperature profiles of \emph{R--T} results are accurate. Thus, whatever anomaly the PnM-Nb sample exhibits, any criticism arising from thermometry is invalid.

In the first cooling process (see bottom panel of \cref{fig4}(a)), the PnM-Nb sample undergoes the SC transition at 9 K, the usual $T_{c}$ for Nb, and returns to the normal state at the same $T_{c}$ in the subsequent warming process. During the first temperature cycle, the PnM-Nb sample thus exhibits usual SC properties. In the subsequent temperature cycle, by contrast, the sample begins to exhibit drastic changes. In the second cooling process (blue curve), the $R$ abruptly drops at 175 K. Because of the two-probe method adopted for this measurement, a tiny resistance residing in electrical pads ($\lesssim 0.2~\si{\ohm}$) remains in the $R$. The residual resistance aside, the resistance of the PnM-Nb sample part, i.e. Nb holey sheet, decreases by at least three orders of magnitude, giving evidence for the Nb holey sheet being superconducting below 175 K. In the subsequent warming process (red curve), the zero-resistance state keeps going through 175 K, and a finite resistance appears at 290 K.

One could, of course, speculate that the abrupt decrease in resistance at 175 K upon cooling is due to an electrical issue with measuring the sample. Indeed, weak points in the suspended holey structure, e.g. narrow bridge-like constrictions (see \cref{fig2}), might break after several thermal cycles because of thermal contractions. However, if some of the conducting paths between electrical pads abruptly open, the two-probe resistance would increase, not dropping to zero, as explained in Ref.~\cite{Uraz2019}. Additionally, such an electrical issue hardly explains why a finite resistance appears again in the subsequent warming process since that would imply that the supposedly opened conducting paths were connected again spontaneously by some means. Thus, such a scenario assuming electrical issues with measuring samples cannot explain either the zero resistance at high temperatures or the phase transition with thermal hysteresis; an alternative scenario is necessary.

Uninterruptedly, \emph{R--T} measurements were repeatedly performed. The result is shown in \cref{fig4}(b). The resistance drop, which was very sharp in the second temperature cycle, is obviously broadened by increasing the number of temperature cycles. The broadening of resistive transition is often observed when magnetic flux is involved in the SC transition, being typical for the so-called type-II superconductors. The more magnetic flux penetrating the sample, the wider the resistive transition becomes. Therefore, if the subject of this study is superconductivity, the experimental fact of resistive broadening indicates that the number of magnetic flux penetrating the sample is increased by repeating \emph{R--T} cycles.

The involvement of magnetic flux upon superconducting transition, despite the absence of an externally applied magnetic field, is a natural consequence of the theory of \emph{hole superconductivity} which describes the dynamics of superconducting transition. In Ref.~\cite{HirschWire2021}, the theory is applied to a wire conducting an electric current, and the dynamics of its self field is provided. Now, that situation is applied to this study, where an Nb holey sheet carries an excitation current for the \emph{R--T} measurement, as shown in \cref{fig4}(c). The theory of hole superconductivity predicts a significance of conduction through anions for the discovery of new high-$T_{c}$ superconductors~\cite{Hirsch1989}, and the SEM-EDX image of the Nb holey sheet has proven that narrow bridge parts contain high amounts of oxygen (\cref{fig3}(b)). Hence, we focus on the conduction through narrow bridges and discuss its superconducting transition. In the normal state (upper panel), an excitation current flows parallel to a bridge, and it generates a self magnetic field in the interior of the bridge. Upon superconducting transition, the excitation current becomes a surface current that can only flow near the surface of a bridge (bottom panel). Because the motion of magnetic field lines is intimately tied to the motion of charge carriers (as proposed in the theory of hole superconductivity), the self field is expelled from the interior of a bridge. It is difficult to observe this effect clearly in conventional metals becoming superconducting, because they are aggregates of three-dimensional submicron grains, and the expelled flux would be averaged out among the grains with random size and random orientation. For a highly oriented system such as the one in this study, by contrast, the effect is significant, and the expelled flux would be trapped in the adjacent large void having the diameter of 20 $\si{\micro m}$. By taking into account the heterogeneity of oxygen concentrations in bridges (see SEM-EDX image in \cref{fig3}), some of them would become superconducting and the others would not. In such a percolative superconducting network with low spatial symmetry, it is unlikely that the magnetic orientation of the trapped flux within the voids would be completely canceled out.

To conclude, each time the Nb holey sheet carrying current undergoes the superconducting transition of percolative nature, the self field is excluded from the body and almost concurrently trapped in the adjacent voids. In fact, even in the absence of an applied field, the width of its resistive transition broadens as the number of temperature cycles increases, as confirmed in \cref{fig4}(b). Further evidence for flux trapping in the holey sheet can be found in magnetic results shown in the next sections.

The superconducting transition, which evolves with magnetic flux, can be a first order phase transition showing hysteresis~\cite{Faber1952,Tera2013}. And the most important feature of such a resistive transition with thermal hysteresis is that a rise of resistance upon warming occurs \emph{always} at a \emph{higher} temperature than the drop of resistance upon cooling. The fact that the high-$T_{c}$ transitions for \emph{ALL} the \emph{R--T} cycles shown in \cref{fig4} share that key feature specific to a first-order phase transition validates this exotic kind of superconducting transition. The reason why warming is always beneficial to superconductivity of this kind can be found in the caption of \cref{fig4}(d), from Ref.~\cite{HMarxiv2021Jan}, using the sixth-order Ginzburg-Landau (GL) free energy density.

\subsection{\emph{M--T} results}
\label{sec:meiss}
\begin{figure}
\centering
\includegraphics[width=1.0\linewidth]{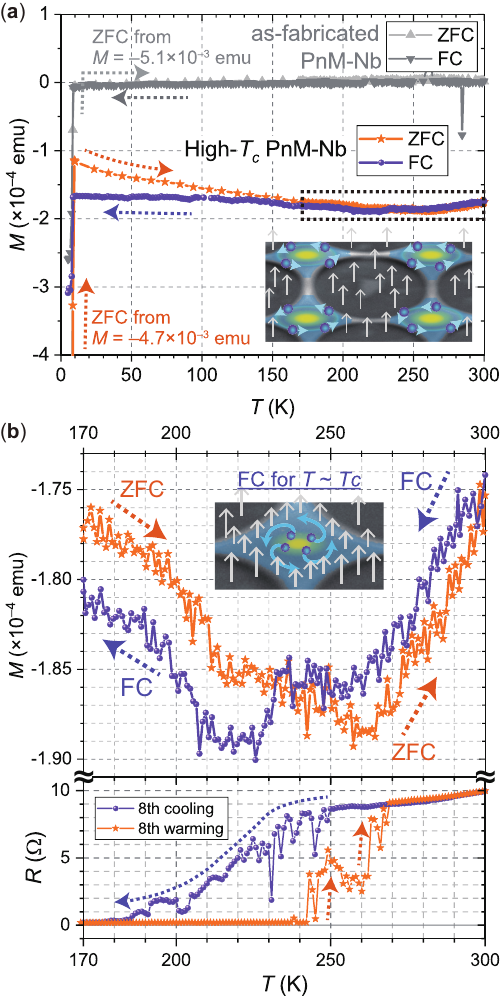}
\caption{\textbf{\emph{M--T} results.} (\textbf{a}) Magnetization versus temperature of an as-fabricated PnM-Nb sample (gray colors) and that of the PnM-Nb sample that exhibited high $T_{c}$'s in the preceding \emph{R--T} measurement (orange and purple colors). An applied magnetic field for both the \emph{M--T} measurements was 1000 Oe, perpendicularly to the sample surface. (Inset) Schematic illustration of the PnM-Nb sample during the ZFC process for $T<T_{c}$ using a false-color SEM. Out-of-plane gray arrows, applied magnetic flux. In-plane cyan arrows, circulating negative charge particles in the counterclockwise direction in such a way as to shield the interior of superconducting islands from the applied magnetic flux. (\textbf{b}) Enlarged \emph{M--T} result in the temperature range of 170--300 K, surrounded by the black dotted square in (a). (Bottom panel) Duplicated \emph{R--T} result of the eighth cycle in \cref{fig4}(b) which was performed just before this \emph{M--T} measurement. (Inset) Schematic illustration of the magnetic flux expulsion during the FC process, the Meissner effect.
\label{fig5}}
\end{figure}

After the eighth \emph{R--T} cycle, the sample chip was taken out of the PPMS, and its temperature ($T$) dependence of magnetization ($M$) was investigated using the MPMS under a perpendicular magnetic field of 1000 Oe, as shown in \cref{fig5}(a). First, a ZFC measurement was performed while warming the sample chip from 4.2 K to 300 K (orange). Subsequently, an FC measurement was performed while cooling it from 300 K to 5.1 K (purple). For comparison, another PnM-Nb sample chip, to which the \emph{R--T} procedure had not been applied yet and therefore that should not have zero resistance yet, was also measured (gray curves). Both sample chips have the same structural geometry. The only difference is whether the microprocessed part of the sample chip, i.e. Nb holey sheet, is in the anomalous zero-resistance state or not.

As the temperature was increased from 4.2 K, the $M$ of the as-fabricated sample (gray) increased from $-5.1\times 10^{-3}$ emu to an approximate zero value of the order of $10^{-6}$ emu at 9 K, indicating an overall disappearance of superconducting diamagnetism of Nb constituting the sample. By contrast, the $M$ of the sample under study that contains the anomalous region (orange) does not reach zero at 9 K, remaining negative. As indicated by the \emph{M--T} curve above 9 K for the reference sample (gray), both diamagnetism of the bulk Si substrate and paramagnetism of the entire Nb film surrounding the microprocessed part (see \cref{fig2}) do not contribute to the magnitude of $M$ any more than the order of $10^{-6}$ emu. Hence, the anomalous behavior in $M$ above 9 K (orange) should be attributed to the microprocessed part with anomalous zero resistance. The magnitude of the remaining $M$ ($-1.1\times 10^{-4}$ emu) is not trivial at all since the MPMS has the sensitivity of $10^{-8}$ emu. The remaining $M$ is approximately 50 times smaller than the diamagnetism below 9 K ($-4.7\times 10^{-3}$ emu), because the area of the Nb holey sheet is only 0.3$\times$0.5 mm\textsuperscript{2}, two-orders-of-magnitude smaller than that of the surrounding Nb film that yielded the perfect diamagnetism below 9 K.

As the temperature is increased from 9 K, the ZFC value (orange) decreases gradually from $M\approx -1.1\times 10^{-4}$ emu to more negative values. When considering the \emph{R--T} result, the sample in this temperature range is in the superconducting (SC) state. Therefore, the \emph{M--T} curve should show a flat temperature dependence if there were no magnetization other than SC diamagnetism. A possible reason for the discrepancy is flux trapping. As shown in a false-color SEM in the inset, there is a non-material part in the PnM sample---the void. Because of its large diameter of approximately 20 $\si{\micro m}$, applied flux easily invades the void. Once it invades, it remains trapped and moves together with the sample. That is, the SQUID ring detects an extra magnetization in addition to the SC diamagnetism residing in the material part. Since the direction of the extra magnetization is parallel to the applied field, the value of the measured $M$ increases, concealing the flatness of the temperature independent SC diamagnetism.

However, the extra magnetization owing to flux trapping is unfavorable to thermodynamic equilibration. As shown soon, the critical field $H_{c}$ for the PnM sample is very large. Because of its significantly large $H_{c}$, the thermodynamic equilibrium state of this sample under the field of 1000 Oe during this \emph{M--T} measurement is not the intermediate state but the perfect shielding state. For such a superconducting sample, the extra magnetization residing in the void is nothing but an unwanted source of thermodynamic nonequilibration. In other words, the extra magnetization owing to flux trapping decreases its magnitude as the temperature is increased, and the ZFC curve exhibits thus monotonically decreasing behavior.

As the temperature is increased further to 300 K, by contrast, the ZFC curve stops decreasing and alters its trend upward. The \emph{M--T} result in the temperature range of 170--300 K is enlarged on the upper panel of \cref{fig5}(b). The lower panel shows the eighth \emph{R--T} result in the same temperature range duplicated from \cref{fig4}(b). (The \emph{R--T} measurement was performed just before this \emph{M--T} measurement.) Precisely describing the complicated dynamics is difficult at this time. Yet it is remarkable that the $M$ measured in the ZFC ``warming'' process (orange) is flipping its trend upward in the temperature range of 250--270 K which is the same temperature range where the resistance $R$ started to rise in the \emph{R--T} ``warming'' process (orange). Based on the assumption that the appearance of $R$ is due to disappearance of superconductivity, it would be reasonable to conclude that the flip of $M$ across 250--270 K is the indication that the SC diamagnetism residing in the PnM material part is disappearing or at least weakening.

Uninterruptedly, the \emph{M--T} measurement was continued while ``cooling'' the sample from 300 K, i.e., the FC measurement was performed. As the temperature is lowered, the FC value (purple) decreases. The lowering of $M$ indicates that the applied magnetic flux is expelled from the interior of an examined specimen, and so far it cannot be accounted for by any physics other than superconductivity, the Meissner effect. The FC value reaches its minimum at the temperature below 230 K which is roughly consistent with the temperature range where the resistance $R$ started to vanish in the \emph{R--T} ``cooling'' process (purple).

The FC curve (purple) flips its trend upward below 210 K, which would be attributed to flux trapping in a non-material part, void. The flux trapping takes place after the PnM material part undergoes a superconducting transition. As the direction of the trapped flux is parallel to the applied field, the value of $M$ thus increases. Below 170 K, however, the FC curve exhibits a relatively flat temperature dependence in contrast to the ZFC curve (see orange and purple curves in \cref{fig5}(a)). For the ZFC measurement, the sample was already cold before the field was applied and, of course, did not know whether the field would be applied or not. Therefore, there was no choice for the sample other than to admit the extra magnetization in the voids. For the FC measurement, by contrast, the sample was cooled under the existence of the applied field. That is, there was a chance for the emergent superconducting screening currents to draw the most suitable geometric pattern that nullifies not only the field invading into the interior of the material part but also the other one trying to reside in voids. Hence, the thermodynamically unfavorable magnetization arising from voids was weakened during the FC measurement to the extent possible, and the FC curve exhibits thus relatively flat temperature dependence.

Finally, it is noteworthy that the value of $M$ still remains negative at 300 K, approximately $-1.75\times 10^{-4}$ emu. Again, the magnitude is significant; such a large negative $M$ cannot be attributed to a bulk Si substrate or an Nb film covering the sample chip whose magnetization are only of the order of $10^{-6}$ emu. Magnetic impurity mixing as a source of the large negative $M$ is also impossible, because such a simple mixing cannot make the clear separation between the ZFC and FC curves observed below 170 K (see orange and purple curves in \cref{fig5}(a)). Hence, there is no alternative but to suppose that the observed large negative $M$ at 300 K is attributed to the SC diamagnetism of the Nb holey sheet. Because the minimum $M$, approximately $-1.89\times 10^{-4}$ emu, is reached at lower temperatures, it is reasonable to assume that not whole but just some part of the holey sheet is in the SC state at 300 K. This assumption is consistent with the percolative superconductivity discussed in the previous section, that the holey sheet as an anion network is composed of a mixture of superconducting and nonsuperconducting paths. It also explains why the resistance at 300 K was lowered from 28 $\si{\ohm}$ to a several $\si{\ohm}$ after the manifestation of high-$T_{c}$ superconductivity (\cref{fig4}(b)). Anyway, if really superconductivity survives at 300 K, then trapped flux should still be detectable at 300 K.

\subsection{\emph{M--H} results at 300 K}
\label{sec:magflutra}
\begin{figure}[h!]
\centering
\includegraphics[width=0.99\linewidth]{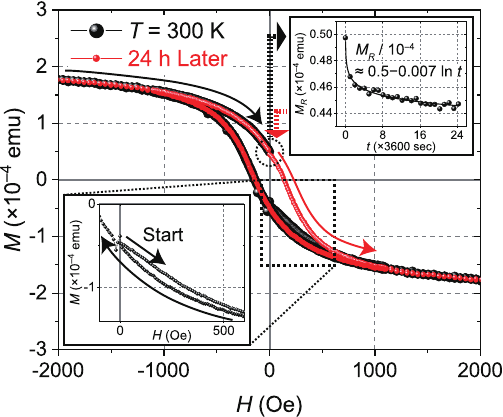}
\caption{\textbf{\emph{M--H} results at 300 K.} Magnetic field dependence of magnetization of the PnM-Nb sample (black color) and that measured again 24 hours later (red color). Both were measured at 300 K. (Lower-left inset) Enlarged \emph{M--H} result nearby the origin surrounded by the black dotted square. (Upper-right inset) Time evolution of the remnant magnetization measured at 300 K for 24 hours under zero field.
\label{fig6}}
\end{figure}

After the \emph{M--T} measurement, its magnetic field ($H$) dependence of magnetization ($M$) was measured at 300 K. As shown in the lower-left inset of \cref{fig6}, the \emph{M--H} measurement was started with the applied field increased from 0 Oe. It is remarkable that at $H=$ 0 Oe the $M$ already shows a non-zero value, approximately $-0.5\times 10^{-4}$ emu. After the FC measurement performed in advance of this \emph{M--H} measurement, the temperature was raised to 300 K with the applied field of 1000 Oe unchanged, then, after the temperature reached 300 K, the applied field was decreased to 0 Oe. Meanwhile, the decreasing magnetic field should reduce the magnitude of trapped flux in voids. Then, if really some part of the Nb holey sheet was still in the SC state, supercurrents flowing in the holey sheet should circulate around the voids in such an inductive way as to prevent the trapped flux from decreasing its magnitude. In other words, in a decreasing field a supercurrent circulating around a void must increase its magnitude. By taking into account the fact that such a circulating supercurrent consists of negative charge particles and by taking a look at the schematic illustration in the inset of \cref{fig5}(a), it can be found that the tangential direction of such circulating negative charge particles around the void is the same as that of negative charge particles circulating around the periphery of superconducting islands generating SC diamagnetism. That is, the SC diamagnetism is enhanced in a decreasing field, therefore, even when the field is decreased to zero, the value of $M$ still remains negative as shown at the beginning of this \emph{M--H} measurement.

As the virgin curve shows (lower-left inset of \cref{fig6}), the SC diamagnetism increases with $H$ increased. The low-field diamagnetism is a usual behavior of a superconductor, but the monotonic behavior is lasting even at $H=$ 2000 Oe. As shown soon, the critical field $H_{c}$ for the PnM sample is very large, and the 2000 Oe turns out to be indeed low enough when compared to the $H_{c}$.

When $H$ is subsequently decreased from 2000 Oe, the SC diamagnetism decreases, thus $M$ increases. It is remarkable that the hysteresis behavior begins at 1000 Oe that is the same magnitude of the magnetic filed applied to the PnM sample during the preceding \emph{M--T} measurement, implying that the flux indeed remain trapped in voids. The decreasing $H$ tries to reduce the magnitude of trapped flux in the array of voids. Due to the same physics explained above, in the decreasing field the SC diamagnetism is enhanced, thus, the measured $M$ becomes lower when $H$ is decreased versus when it is increased as can be seen in the lower-left inset. This is a universal feature of thermomagnetic hysteresis caused by flux trapping. For a good example, see Fig.~2 in Ref.~\cite{HirschGranular2022}.

When $H$ is increased from $-2000$ Oe and reaches 0 Oe (see main panel of \cref{fig6}), the PnM sample exhibits a positive $M$, approximately $0.5\times 10^{-4}$ emu. This is a remnant magnetization. At 300 K and in the absence of an applied field, the time ($t$) evolution of the remnant magnetization ($M_{R}$) was measured for 24 hours. The result is shown in the upper-right inset. At finite temperatures, thermal energy may allow flux lines to jump from one void to another in response to flux-density gradient, hence, there may be an observable decrease of the magnitude of trapped flux with time~\cite{TinkhamBook}. Since any creep will relieve the gradient, the creep gets slower and slower. That is, the time dependence is logarithmic, and so is the observed \emph{$M_{R}$--t} curve. Given that the driving force is proportional to the magnitude of trapped flux, the exponential dependence of the creep rate on the driving force is
\begin{equation}
\frac{dM_{R}}{dt}\propto -e^{M_{R}/C_{1}},\label{eq:Mr}
\end{equation}
which has the solution $M_{R}=C_{2}-C_{1}\ln t$, where the constants $C_{1}$ and $C_{2}$ can be obtained from the fitting curve shown in the figure. Then it is possible to estimate how long the trapped flux persists in the array of voids and hence how long the circulating supercurrents in the holey sheet will take to die out at 300 K. It will take $t=e^{C_{2}/C_{1}}\approx e^{0.5/0.007}\approx 10^{31}$ s $\approx 10^{23}$ years.

When compared to the pioneering study for a tubular sample made of hard-superconductor NbZr measured at 4.2 K~\cite{Kim1962} or to the recent study for a granular superconductor made of graphite powder measured at 300 K~\cite{Esquin2012}, the lifetime is $10^{69}$ or $10^{370}$ times shorter. But still, when considering the thermodynamically unstable nature of trapped flux in the Nb holey sheet, which has a large $H_{c}$ and hence prefers to remain the perfect shielding state, the lifetime of the unfavorable flux trapping is still long, long enough in any practical sense.

After the \emph{$M_{R}$--t} measurement, namely 24 hours later, the \emph{M--H} measurement was performed again with $H$ increased from 0 Oe (red curve in \cref{fig6}). As shown, it retraces the initial hysteresis loop, indicating that the Nb holey sheet is indeed trapping magnetic flux. Whether or not it is due to superconductivity, the result evidences that the holey sheet sustains persistent current at 300 K. Of course, as of now, we do not know of any physics that makes this possible other than superconductivity.

\subsection{Critical field and the Superconducting atom}
\label{sec:hc}
\begin{figure}[h!]
\centering
\includegraphics[width=0.89\linewidth]{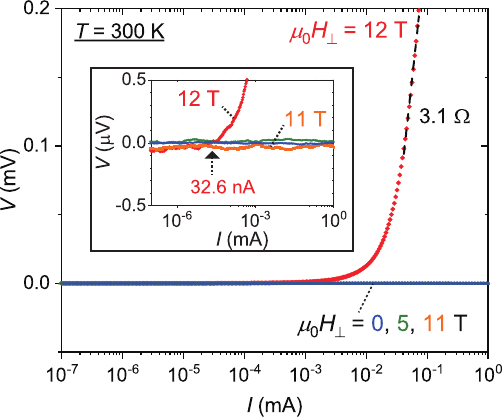}
\caption{\textbf{Critical field at 300 K.} Output voltage versus applied current under various magnitudes of perpendicular magnetic field, measured at 300 K. Inset, zoomed-in plot.
\label{fig7}}
\end{figure}

As mentioned previously, the critical field $H_{c}$ for the Nb holey sheet with zero resistance was investigated at 300 K. Under various perpendicular magnetic fields of $\mu_{0}H_{\perp}=$ 0, 5, 11 and 12 T, the output voltage was measured while applying current. The result is shown in \cref{fig7}. For $\mu_{0}H_{\perp}\leq$ 11 T, the voltage does not increase for applied currents lower than 1 mA. At $\mu_{0}H_{\perp}=$ 12 T, by contrast, a finite voltage is observed for a current exceeding a small value of 32.6 nA as shown in the inset. Hence, the intrinsic $\mu_{0}H_{c}$ at 300 K for the superconducting holey sheet without applied current will be a little bit larger than 12 T.

By taking into account the initial sharp resistance drop (gray curve in \cref{fig4}(b)), the Nb holey sheet is intrinsically the so-called type-I superconductor, therefore, its $H_{c}$ corresponds to the thermodynamic critical field. As explained previously, the large thermodynamic critical field was responsible for the clear separation between the ZFC and FC curves in \cref{fig5}(a). Even if an unknown demagnetizing factor of the holey sample yields a 100 times larger magnitude than that of the actual field applied during the \emph{M--T} measurement (1000 Oe), the product is still lower than the $H_{c}$. That is, the perfect shielding state is indeed favorable for the superconducting holey sample during the \emph{M--T} measurement. In other words, flux trapping in the array of voids is indeed thermodynamically unfavorable. That's why the ZFC curve exhibited the decreasing behavior as the temperature was increased. Also, whenever possible the flux trapping is suppressed as much as possible, that's why the FC cooling curve exhibited the relatively flat temperature dependence. Nevertheless, once trapped under isothermal condition, it remains trapped for a sufficiently long time as proven in the previous section (\cref{fig6}).

By the way, the value of $\mu_{0}H_{c}\gtrsim$ 12 T is much too large when considering the fact that the largest known critical fields for standard type-I superconductors are of order 0.05 T. According to the recent custom, such a potentially disruptive result is to be excluded from the science community~\cite{disnat2023}. And the situation gets worse when such a result cannot be explained by an idol theory such as the BCS theory; unfortunately, that's the case for this study. However, there is an alternative theory that explains the observed extraordinary value, although it was already forgotten a long time ago.

In 1937, following the logical argument raised by Fritz London~\cite{London1937}, John Slater immediately considered a magnetic susceptibility of an atom, and, in order for the atom to have perfect diamagnetism, he gave the minimum radius of the atom,
\begin{equation}
r_{S}=137\times a_{0}~(\approx \mathrm{7.25~nm}),\label{eq:rs}
\end{equation}
where $a_{0}$ is the Bohr radius (0.529 \AA) and the number 137 comes from the fine structure constant ($\alpha\equiv \frac{e^{2}}{\hbar c}\approx \frac{1}{137}$)~\cite{Slater1937}. According to London and Slater, this is the element of any superconductor and cannot be separated into simpler units any further. It is approximately 137 times larger than the Bohr's atom and has perfect diamagnetism with the magnitude corresponding to a single flux quantum, $\phi_{0}$ ($\approx 2.07\times10^{-15}$~Wb). The expanded atom is called the Slater's atom.

Then, when applied flux invade the interior of a superconductor? It is when the applied flux are getting dense, so dense that the Slater's atoms with the radius $r_{S}$ having a diamagnetic $\phi_{0}$ overlap each other. Hence the critical field can be given by
\begin{equation}
\mu_{0}H_{c}=\frac{\rm{\phi_{0}}}{\pi\times r_{S}^{2}}~(\approx \mathrm{12.5~T}).\label{eq:hc}
\end{equation}
The experimentally confirmed extraordinary critical field, which will be a little bit larger than 12 T, is thus explained by the principle of the superconducting atom. This indicates that at least the superconductor presented in this study is composed of the superconducting atoms, as London and Slater expected. And what's most remarkable is its simplicity; both \cref{eq:rs,eq:hc} do not include any single material parameter, implying that the pairing mechanism underlying this superconductivity as well as the forgotten theory are of pure physics. In other words, this study appears to be unrelated to BCS theory despite the initial motivation, because BCS superconductors do function only if material parameters such as an electron--phonon coupling constant $\lambda_{e-ph}$ and a Coulomb pseudopotential $\mu^{*}$ are empirically given~\cite{BCS1957,Mc1968}.

\section{Discussion}
\label{sec:dis}
\subsection{Giant square lattice and the ``\emph{Size}'' of the first-order phase transition}
\label{sec:giant}
The new, apparently superconducting, phase transition \emph{always} exhibited thermal hysteresis, indicating that it is a first order phase transition. This repeatedly confirmed experimental fact is never accounted for by BCS theory, while in 1974, half a century ago, Halperin, Lubensky, and Ma, or HLM for short, concluded that any superconducting transition must \emph{always} be first order because of the intrinsic fluctuating magnetic field~\cite{Halperin1974}. They state, ``\emph{Roughly, the driving force for the first-order transition is the partial expulsion of the `black-body radiation,' (or of the director fluctuations) from the low-temperature phase.}'' This effect becomes significant especially for a highly oriented system with an applied current. In this study, the dynamics of the ``\emph{black-body radiation}'' was described by the theory of hole superconductivity proposed by Hirsch~\cite{HirschWire2021}; a self field generated by an excitation current for measuring resistance is expelled from a body becoming superconducting and almost concurrently trapped in the array of voids. As the dynamics of magnetic flux, one form of the bosonic radiation, is explicitly involved in the superconducting transition, it must be an obvious first-order phase transition with unambiguous hysteresis. In this section, we estimate the ``\emph{size}'' of the first-order phase transition, i.e., the width of the thermal hysteresis, based on the HLM theory.

\begin{figure}[b]
\centering
\includegraphics[width=1.1\linewidth]{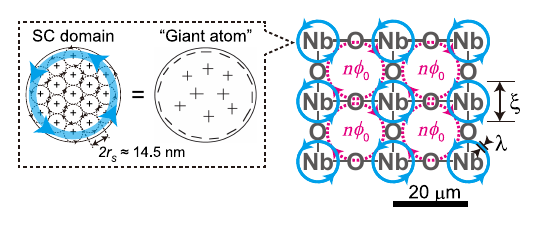}
\caption{(Left panel) Schematic superconducting (SC) domain including a bunch of Slater's atoms with the diameter of $2r_{S}\approx 14.5~\si{nm}$, and its superpositioned equivalent as a ``giant atom'' reprinted from the theory of hole superconductivity, Ref.~\cite{HirschPLA2001}. ``$+$'' and ``$-$'' indicate a positive and negative charge, respectively. Cyan arrow, a superconducting screening current. (Right panel) Schematic representation of the possible room temperature superconductor under study. ``Nb giant atoms'' form a ``giant square lattice'' networked by oxygen.{\if0 Cyan arrows, a supercurrent generated at the moment when electrons go superconducting. Magenta arrows, the dual of the supercurrent. It circulates in the clockwise direction, thus producing $n\phi_{0}$ ($n=$ 1, 2, ...) in its interior, i.e., in the void of the giant square lattice.\fi} Under this picture, the superconducting coherence length $\xi$ and the penetration depth $\lambda$ roughly correspond to the diameter of an Nb giant atom and the diameter of a Slater's atom, respectively, which are approximately 15 $\si{\micro m}$ and 14.5 nm.
\label{fig8}}
\end{figure}

From the experimental results for the critical field, we considered that superconducting (SC) domains, which are the building blocks of the Nb holey sheet, are composed of Slater's atoms. It is noteworthy that the Hirsch's theory being independent of the London and Slater's argument reaches the same consequence~\cite{HirschPLA2001}, as shown in the left panel of \cref{fig8}. In the SC domain, Slater's atoms with the diameter of $2r_{S}$ are arranged in such a way as to be associated with \cref{eq:hc}. In the inner domain, the superposition of electrons orbits nullifies themselves, leaving electrons orbits only within a thin layer adjacent to the periphery. By the superposition, only positive charges, i.e. holes, remain in the bulk, and negative charges are globally expelled from the bulk to the periphery. Indeed it looks like a ``giant atom.'' Under an external magnetic filed, the remaining electrons orbits at the periphery draw a supercurrent path (cyan arrow){\if0 circulating around the SC domain\fi} in such a way as to shield the interior from the applied field. That is, the penetration depth $\lambda$ of an SC domain roughly corresponds to the diameter of a Slater's atom, $2r_{S}\approx 14.5~\si{nm}$.

The right panel of \cref{fig8} shows a schematic representation of the possible room temperature superconductor (RTS) under study. ``Nb giant atoms'' form a ``giant square lattice'' networked by oxygen. By taking into account the definition of the coherence length of superconductivity and the fact that the coherence length of cuprate superconductors roughly corresponds to the lattice constant of a copper--oxygen plane constituting cuprate superconductors, the coherence length $\xi$ of the possible RTS would be in the range of the diameter of an Nb giant atom or the lattice constant of the giant lattice, namely, 15--20~$\si{\micro m}$.

The thus relation, $\xi~(15~\si{\micro m})\gg \lambda~(14.5~\si{nm})$, indicates that the RTS under study is intrinsically a type-I superconductor. In fact, the initial transition to zero resistance was very sharp (see gray curve in \cref{fig4}(b)). In the subsequent \emph{R--T} cycles, on the other hand, the resistive transition became progressively broader with increasing the number of \emph{R--T} cycles, despite zero magnetic field applied during the measurement. This can be considered as follows. The more times the \emph{R--T} cycle is repeated, the more times Nb giant atoms undergo superconducting transition with supercurrents circulated (cyan arrows). And the more times supercurrents circulate around the Nb giant atoms, the more times the dual of the supercurrents (magenta arrows) circulate around the voids within the giant lattice. Thus, the more $\phi_{0}$'s the giant lattice produces and traps in its voids despite zero applied field. The trapping of $\phi_{0}$'s after multiple \emph{R--T} cycles changes the giant lattice of type-I nature into the so-called type-II superconductor with the resistive transition broadened. The argument here is valid only for ideal two-dimensional superconductors with high orientation and does not apply to three-dimensional aggregation of submicron grains or three-dimensional stacking of two-dimensional planes where high orientation is not expected. In other words, such an unambiguous type change has never been manifested in any superconductor before, conventional or unconventional.

Therefore, the superconducting giant lattice could potentially be the answer to the HLM's following question: ``\emph{It is not clear whether it is possible to prepare a sample with the homogeneity necessary to observe the first-order transition.}'' According to the HLM theory~\cite{Halperin1974}, the width of thermal hysteresis is given by
\begin{equation}
\Delta T_{1}\propto \kappa^{-6},\label{eq:hlm}
\end{equation}
where $\kappa$ is the ratio of the London penetration depth to the coherence length in the superconducting state. For the best known type-I materials, $\kappa$ of order 0.02 was used, and the ``\emph{size}'' of the first-order transition $\Delta T_{1}$ was estimated to be at most a few $\si{\micro K}$. For the giant lattice under study, by contrast, $\kappa$ ($\equiv \lambda/\xi$) is of order 0.001. Therefore, $\Delta T_{1}$ can readily reach the order of 100 K, and actually it did so in the \emph{R--T} measurement (see blue and red curves in \cref{fig4}(a)).

It should be noted that \cref{eq:hlm} is valid for spatial three-dimensional systems. Because the holey sheet under study is a spatial three-dimensional system with the thickness of 150 nm, \cref{eq:hlm} is applicable, and the above argument about the ``\emph{size}'' of the first-order phase transition is correct. Since it is a spatial three-dimensional system, a long-range order can be developed according to Mermin-Wagner theorem, and two-dimensional superconductivity has been manifested in this study. In fact, \cref{eq:hc}, which gives the critical field for this superconductor, is obtained by assuming that Slater's atoms are arranged in two dimensions. Although a later study of a lattice model for superconductivity in three dimensions by Dasgupta and Halperin in 1981 made a strong case that the SC transition in three dimensions can be second order~\cite{Halperin1981}, it of course does not apply to the two-dimensional superconductivity of this study.

The value of $\lambda$ and the extraordinarily large $\xi$ will be confirmed experimentally in the forthcoming paper by studying the giant lattice as a Josephson junction array.

\subsection{Other considerations for mass production of the possible room temperature superconductor}
\label{sec:manu}
To date, 18 PnM-Nb samples in a rectangular shape were examined, and 7 samples exhibited zero resistance at high temperatures (see Supplementary material). So the yield rate is 39\%. All the samples retained the zero resistance at higher temperatures in a warming process than in a cooling one, as with the \emph{R--T} results shown in \cref{fig4}. ``\emph{The benefit of warming to superconductivity}'' sounds anomalous but is a fundamental feature of superconducting transitions with thermal hysteresis. The fact that \emph{all} the superconducting transitions presented in this study, including those in the Supplementary material, share this significant feature rather guarantees such exotic superconducting transitions. In addition, the zero resistance was attainable regardless of the sample size and regardless of whether the resistance was measured by the two or four probe method. However, the $T_{c}$ varied with the sample, taking some typical values: 35 K, 50 K, 100 K and 175 K. Additionally the number of \emph{R--T} cycles repeatedly applied to the sample in order to change it into the zero-resistance state varied with the sample. A possible reason for those varieties and an effective way to enhance the yield rate are discussed below.

\begin{figure}[h!]
\centering
\includegraphics[width=0.72\linewidth]{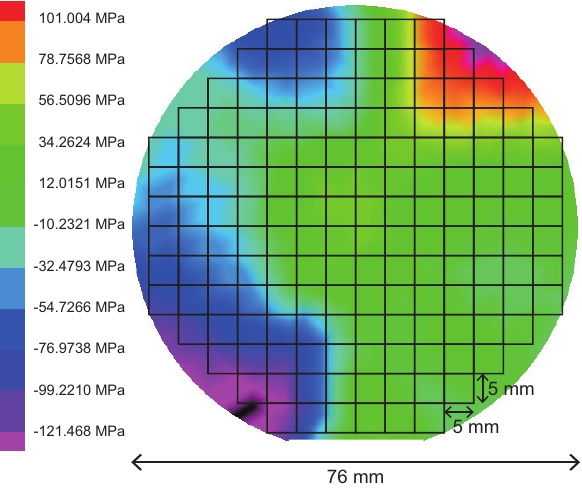}
\caption{In-plane stress of an Nb layer sputtered on an SiO\textsubscript{2}/Si wafer with the diameter of 76 mm. The size of a sample chip (5$\times$5 mm\textsuperscript{2}) is also indicated. The in-plane stress was measured using FSM 128NT (Frontier Semiconductor).
\label{fig9}}
\end{figure}

As written in \cref{sec:samfab}, those samples were microfabricated on a 3-inch ($\approx$ 76-mm) wafer, and each sample chip with the size of 5-mm squares was taken from the wafer. As shown in \cref{fig9}, there is a distribution of in-plane stress in the Nb layer deposited on the SiO\textsubscript{2}/Si wafer. In-plane stress is widely distributed in the range from $-$120 to $+$100 MPa. That is, each sample chip has a different in-plane stress to each other, and the different in-plane stress may yield a different magnitude of lattice expansion after the final HF dry etching process, as indicated by XRD spectra (\cref{fig3}(a)). Also, the different magnitude of lattice expansion may affect oxidation of the self-standing PnM sample (\cref{fig3}(b)). Therefore, if the high-$T_{c}$ zero-resistance state found in this study is truly promoted by the square-lattice oxygen network as is the case for cuprate high-$T_{c}$ superconductors, the different magnitudes of lattice expansion and oxidation may be highly important factors in determining whether the thus microfabricated PnM sample undergoes the extraordinary transition or not. Hence, controlling in-plane stress and oxygenation during the microfabrication process will be the next important step to increase the yield rate for the zero-resistance state. The former can be done by adjusting pressure and flow rate of Ar gas during the Nb sputtering deposition. Current oxygenation, on the other hand, depends on natural oxidation. An alternative method, for example applying oxygen plasma ashing to PnM samples prior to \emph{R--T} cycles, will be better to control oxidation.{\if0 Investigating how they affect \emph{R--T} results will be the most significant study.\fi}

Since the zero-resistance state presented in this study was for the first time observed in March 2018, many \emph{R--T} measurements had been performed, but none of them had succeeded in reproducing the zero-resistance state. The aforementioned yield rate (39\%) is achieved after an inevitable condition during the \emph{R--T} cycles was found in November 2018, namely the slow temperature sweep rate. A typical \emph{R--T} recipe is summarized in \Cref{tab1}. The necessity of the slow rate especially for the temperature range of 20--60 K is not well understood yet but may be related to absorbed oxygen in the PnM sample since oxygen is known to undergo an antiferromagnetic transition below 54 K and to change its solid structure ($\beta \leftrightarrows\gamma$ phase) at 43 K. At least, we already know that the holey sheet contains high amounts of oxygen in its bridge-like parts (see SEM-EDX image in \cref{fig3}). The relevance of the oxygen structural transition to the zero-resistance state has to be investigated together with the enhancement of the yield rate. Preceding studies on the correlation between ionic displacements and electronic motion in a copper--oxygen plane~\cite{JT2008} or others on the anion ordering in an organic superconductor (TMTSF)\textsubscript{2}ClO\textsubscript{4}~\cite{Pouget1990}, for instance, will be useful to start on the investigation. The latter is particularly interesting because its superconducting phase appears only when it is cooled slowly through the structural transition temperature for \ce{ClO4^-} anions~\cite{Haddad2007,Yone2018}, sharing the same cooling condition under which the holey sheet in this study as an \ce{O^-} anion network develops superconductivity.
\begin{table}[h!]
\caption{\label{tab1}~Typical \emph{R--T} recipe$^*$ for the zero-resistance state}
\vspace{-4mm}
\begin{center}
\footnotesize
\begin{tabular}{@{}llcc}
\toprule
\begin{tabular}{l}\emph{R--T} cycle\\number\end{tabular}&\begin{tabular}{c}Temperature sweep rate\end{tabular}&\begin{tabular}{c}Temperature\\approaching\\mode$^{\dag}$\end{tabular}&\begin{tabular}{c}Elapsed\\time\end{tabular}\\
\toprule
1st~--~6th&300~$\rightarrow$~~~60 K: 1.0 K/minute&S&\\
&~~60~$\rightarrow$~~~20 K: 0.5 K/minute&F&\\
&~~20~$\rightarrow$~~~~~2 K: 1.0 K/minute&S&\\
&~~~~2~$\rightarrow$~~~20 K: 1.0 K/minute&S&\\
&~~20~$\rightarrow$~~~60 K: 0.5 K/minute&F&16~$\sim$~18\\
&~~60~$\rightarrow$~300 K: 1.0 K/minute&S&hours/cycle\\
\midrule
7th to&300~$\rightarrow$~~~~~2 K: 1.0 K/minute&S&9~$\sim$~10\\
the last&~~~~2~$\rightarrow$~300 K: 1.0 K/minute&S&hours/cycle\\
\bottomrule
\end{tabular}
\end{center}
\vspace{-2mm}
\footnotesize{
$^{*}$This recipe is based on the assumption that the PPMS is used with its protocol described in \cref{sec:meamet}. Of course it is possible to find a better recipe that enables a quicker realization of the zero-resistance state within the smaller number of \emph{R--T} cycles.
\\[2pt]
$^{\dag}$Letter ``S'' and ``F'' indicate that the resistance is measured while sweeping the temperature and fixing it at each measurement point, respectively.
}
\end{table}

\section{Conclusion}
\label{sec:conc}
Modification of the phonon system of a conventional superconductor did not change its $T_{c}$ in the temperature range for conventional superconductors, shaking our confidence in the BCS theory. Rather, the engineered metallic sheet unexpectedly formed a partly oxygenated square lattice, reminiscent of a copper--oxygen plane in unconventional cuprate superconductors with a high $T_{c}$. Its magnetization decreased near room temperature, the same temperature at which its resistance dropped to zero, showing a feature associated with an onset of superconductivity, the Meissner effect. Also, a remnant magnetization was clearly detected at 300 K in the absence of an applied magnetic field; according to M\"{u}ller et al.~\cite{Muller1987}, ``\emph{it is a proof of superconductivity in its own right.}'' Thus, this study meets established criteria for a conclusive demonstration of true superconductivity~\cite{Sridhar2023}, namely: (I) True zero resistance demonstrated by persistent currents and (II) The Meissner effect.{\if0 This study began with the BCS theory. However, that beautiful theory does not agree with the experiments. (You are expected to recall the Feynman's famous quote: ``\emph{It doesn't matter how beautiful your theory is, it doesn't matter how smart you are. If it doesn't agree with experiment, it's ... .}'') Thus, this study provides an opportunity to reevaluate the validity of the established BCS theory.\fi} The macroscopic quantum perspective on superconductivity with an energy gap had already been introduced by London and Slater before BCS, as a \emph{big superconducting atom}~\cite{London1937}. The prediction that the essence of superconducting transition lies in a \emph{first order phase transition}{\if0, which may currently be difficult for everyone to accept,\fi} was already made by Halperin, Lubensky, and Ma~\cite{Halperin1974}. The dynamics of superconducting transition, namely, the formation of the big superconducting atoms in a material, has already been provided by \emph{the theory of hole superconductivity} proposed by Hirsch~\cite{HirschPLA2001}. It appears reasonable to explore further in depth based on these clues.
\\
\\
\noindent
{\small
{\bf Acknowledgments:}
The author is indebted to Tom Lubensky for pointing out that the HLM theory is valid for spatial three-dimensional systems, and to Yoichi Higashi for helpful discussions regarding the difference between spatial dimension and order parameter dimension. He also wishes to thank Jim Ashburn for discussing how the monumental high-$T_{c}$ superconductor YBCO was discovered~\cite{Ashburn1987,JimWeb}, and to Masato Yamawaki for discussing this room temperature superconductor.
\\
{\bf Data availability statement:}
All raw data are available to everyone at an open-access repository~\cite{Zenodo2022}, together with the GDSII file used for sample patterning. And all figures in this paper can be created directly from the raw data without any data processing such as background subtraction.
\\
{}
{\bf Declaration of competing interest:}
The author declares no competing interests.
}
\urlstyle{same}

\end{document}